\documentclass[journal]{IEEEtran}
\usepackage{cite}
\usepackage{hhline}
\usepackage[pdftex]{graphicx}
\ifCLASSINFOpdf
\else
\fi
\usepackage{url}
\begin{document}
\title{Trigger Merging Module\\ for the J-PARC E16 Experiment}
\author{M. Ichikawa, T. N. Takahashi, K. Aoki, S. Ashikaga, E. Hamada, R. Honda, Y. Igarashi, M. Ikeno, D. Kawama,\\
M. Naruki, K. Ozawa, H. Sendai, K. N. Suzuki, M. Tanaka, T. Uchida, and S. Yokkaichi%
\thanks{Manuscript received June 24, 2018. This work was performed by the RIKEN Junior Research Associate Program, a Grant-in-Aid for JSPS Fellows (12J01196), and MEXT/JSPS KAKENHI (Nos. 21105004, 26247048, and 15K17669).}%
\thanks{M. Ichikawa, S. Ashikaga, M. Naruki, and K. N. Suzuki are with Department of Physics, Kyoto University, Kitashirakawa Sakyo-ku, Kyoto 606-8502, Japan.}%
\thanks{M. Ichikawa is with Riken Cluster for Pioneering Research, RIKEN, 2-1 Hirosawa, Wako, Saitama 351-0198, Japan}%
\thanks{T. N. Takahashi is with Research Center for Nuclear Physics (RCNP), Osaka University, 10-1 Mihogaoka, Ibaraki, Osaka, 567-0047, Japan.}%
\thanks{K. Aoki, E. Hamada, Y. Igarashi, M. Ikeno, K. Ozawa, H. Sendai, M. Tanaka, and T. Uchida are with Institute of Particle and Nuclear Studies, KEK, 1-1 Oho, Tsukuba, Ibaraki, 305-0801, Japan}%
\thanks{R. Honda is with Department of Physics, Graduate School of Science, Tohoku University, Sendai, Miyagi 980-8578, Japan}%
\thanks{S. Yokkaichi is with RIKEN Nishina Center, RIKEN, 2-1 Hirosawa, Wako, Saitama 351-0198, Japan}}
\IEEEpubid{\makebox[\columnwidth]{} \hspace{\columnsep}\makebox[\columnwidth]{}}
\maketitle
\begin{abstract}
An experiment to measure an invariant mass of $\phi$ mesons in nuclear medium is planned as the J-PARC E16 experiment.
A trigger merging module (TRG-MRG) has been developed to detect leading-edges from 256 channels of discriminator-output signals and transmit those serialized hit data to trigger decision module with four optical links.
The result of the test shows enough performance of the TRG-MRG as 1\,ns TDC and data multiplexer with four 6.25\,Gbps transceivers.
\end{abstract}
\IEEEpeerreviewmaketitle
\section{Introduction}
\IEEEPARstart{A}{} major part of the hadron mass in vacuum is considered to be originated from the spontaneous breaking of chiral symmetry characterized by quark condensate.
It is expected that, at finite density or high temperature, the broken symmetry restores partially and the hadron mass is modified.
The J-PARC E16 experiment measures an invariant mass of $\phi$ mesons in nuclear medium and investigate the partial restoration of breaking of chiral symmetry at nuclear density\cite{E16-1,E16-2}.

In the experiment, $\phi$ mesons are produced in nuclei by the exposure of 30\,GeV proton beam of $1\times10^{10}$\,/pulse, with a duration of 2\,s, to nuclear targets at the high momentum beam-line at J-PARC.
We measure $\phi$ mesons in the electron-positron decay channels and reconstruct the invariant mass.
\subsection{Setup of Spectrometer}
\begin{figure}
\centering
\includegraphics[clip,width=8.5cm]{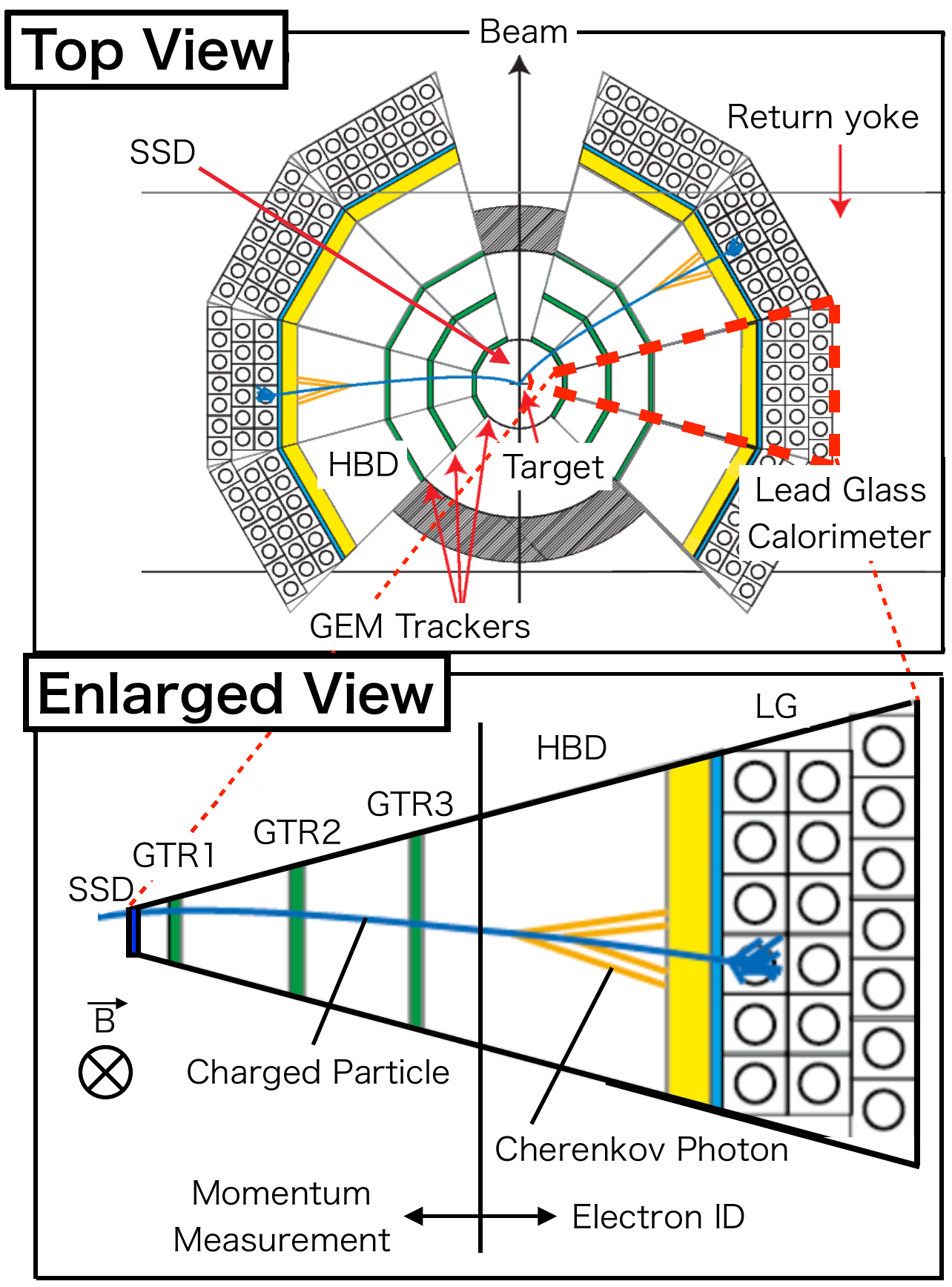}
\caption{(Top) Top view of the spectrometer. (Bottom) Enlarged view of the one-module of detectors.}
\label{spectrometer}
\end{figure}

Figure \ref{spectrometer} shows a top view of the spectrometer.
Four types detectors are used for momentum measurement of electron/positron.
From an inner side, silicon-strip detectors (SSD) and three layers of GEM trackers (GTR1, 2, 3) are located in the strong magnetic field for flight-path detection and momentum reconstruction\cite{GTR}.
The electron/positron identification is performed with hadron-blind detectors (HBD) and lead-glass calorimeters (LG)\cite{HBD}.
The number of readout channels is 112,996 and waveform data from all types of detectors are taken to solve piled up signals.
The waveform data are buffered with modules using APV25-S1 chips\cite{APV} and DRS4 chips\cite{DRS4} with the buffering-time of 4\,$\mu$s and 2\,$\mu$s, respectively\cite{circuit-total}.
Therefore required latency for the trigger signal is less than 2\,$\mu$s.
\IEEEpubidadjcol
\subsection{Trigger System}
\begin{figure}
\centering
\includegraphics[clip,width=8.5cm]{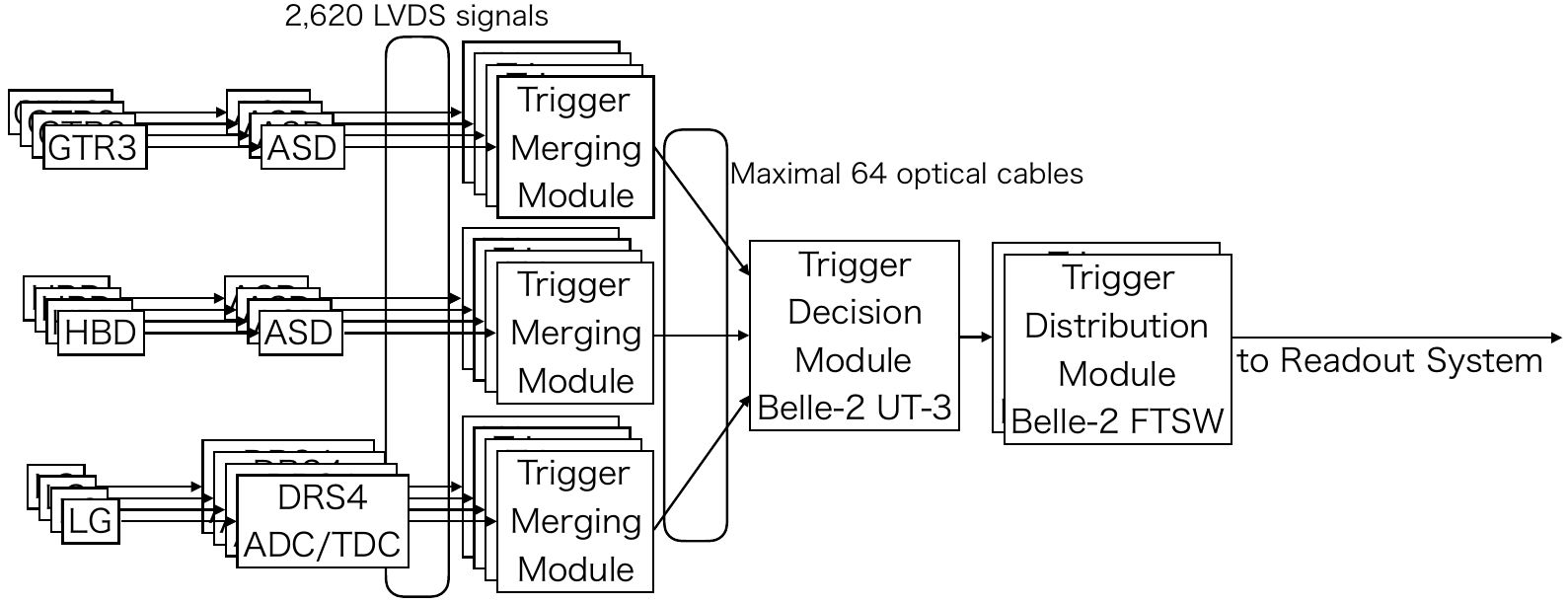}
\caption{Overview of the trigger system.}
\label{trigger-system}
\end{figure}

For the trigger generation, discriminator-output signals from GTR3, HBD, and LG are used.
The number of trigger channels is 2,620.
The maximal single rate is expected to be typically 1\,MHz/ch and the minimal width of the discriminator-output signals is 3\,ns.
Therefore the sampling time must be less than 3\,ns.
Overview of the trigger system is shown in Fig. \ref{trigger-system}.
Analog signals from detectors are discriminated by DRS4 ADC/TDC or ASD (Amplifier-Shaper-Discriminator) developed for the experiment\cite{DRS4-module, GTR-ASD, HBD-ASD}.
In the trigger merging modules, called TRG-MRG, leading-edges are detected and serialized data of them are transmitted to a trigger decision module by optical transceivers.
Belle I\hspace{-1pt}I UT3, that has 16 QSFP+, is used for the trigger decision module\cite{UT3}.
Finally, the trigger signal is distributed to readout modules by Belle I\hspace{-1pt}I FTSW\cite{FTSW}.

The latency before the TRG-MRG is estimated to be 600\,ns, mainly due to the drift time of GTR3.
To make the latency less than the waveform buffering-time of 2\,$\mu$s, we design the latency of the detection of the leading-edge and transmitting, trigger decision, and trigger distribution as 500\,ns, 500\,ns, and 300\,ns, respectively.
With the design value, the total latency including the drift time of GTR3 becomes 1,900\,ns.

The requirement to the TRG-MRG is summarized in TABLE \ref{req-TRG-MRG}.
\begin{table}
\begin{center}
\caption{The requirement to the TRG-MRG.}
\label{req-TRG-MRG}
\begin{tabular}{|l||l|}
\hline
Point&Value \\ \hhline{|=#=|}
Function&TDC + Optical Transceiver \\ \hline
Cable Reduction&2,620 LVDS $\rightarrow$ $<$64 optical cables \\ \hline 
Single Rate&1 MHz/ch \\ \hline
Sampling Time&$<$3 ns \\ \hline
Latency&$<$500 ns \\ \hline
\end{tabular}
\end{center}
\end{table}
\section{Development}
\begin{figure}
\centering
\includegraphics[clip,width=8.5cm]{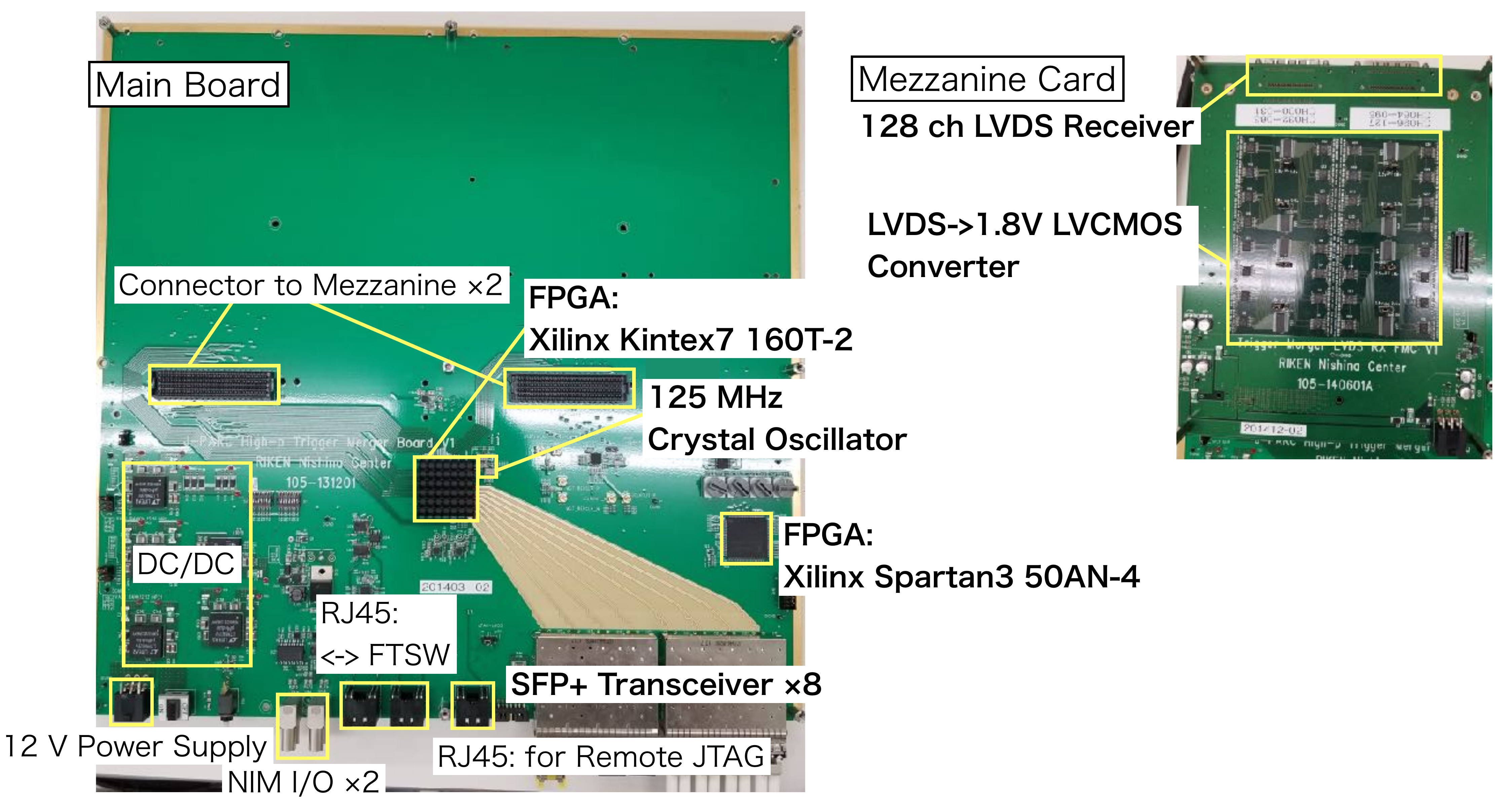}
\caption{The picture of the TRG-MRG.}
\label{TRG-MRG-hard}
\end{figure}

Figure \ref{TRG-MRG-hard} is a picture of the TRG-MRG.
The module consists of one main board and two mezzanine cards.
The mezzanine card has four 32 channels LVDS receivers and converters from LVDS to 1.8\,V LVCMOS format and is replaceable according to the formats of the input connectors.
In the main board, two FPGAs (Xilinx Kintex-7 160T-2 and Xilinx Spartan-3 50AN-4), two crystal oscillators of 125\,MHz, and eight SFP+ transceivers are installed\cite{kintex7, spaltan3}.
The 1\,ns and 256 channels multi-hit TDC and 6.25\,Gbps and four-lanes GTX transceivers are implemented in the Kintex-7 by Vivado2017.2 provided by Xilinx.
The channel reduction from 2,620 channels to maximal 64 optical transceivers is realized by using about 15 TRG-MRGs.
\subsection{Firmware}
\label{firmware}
\begin{figure}
\centering
\includegraphics[clip,width=8.5cm]{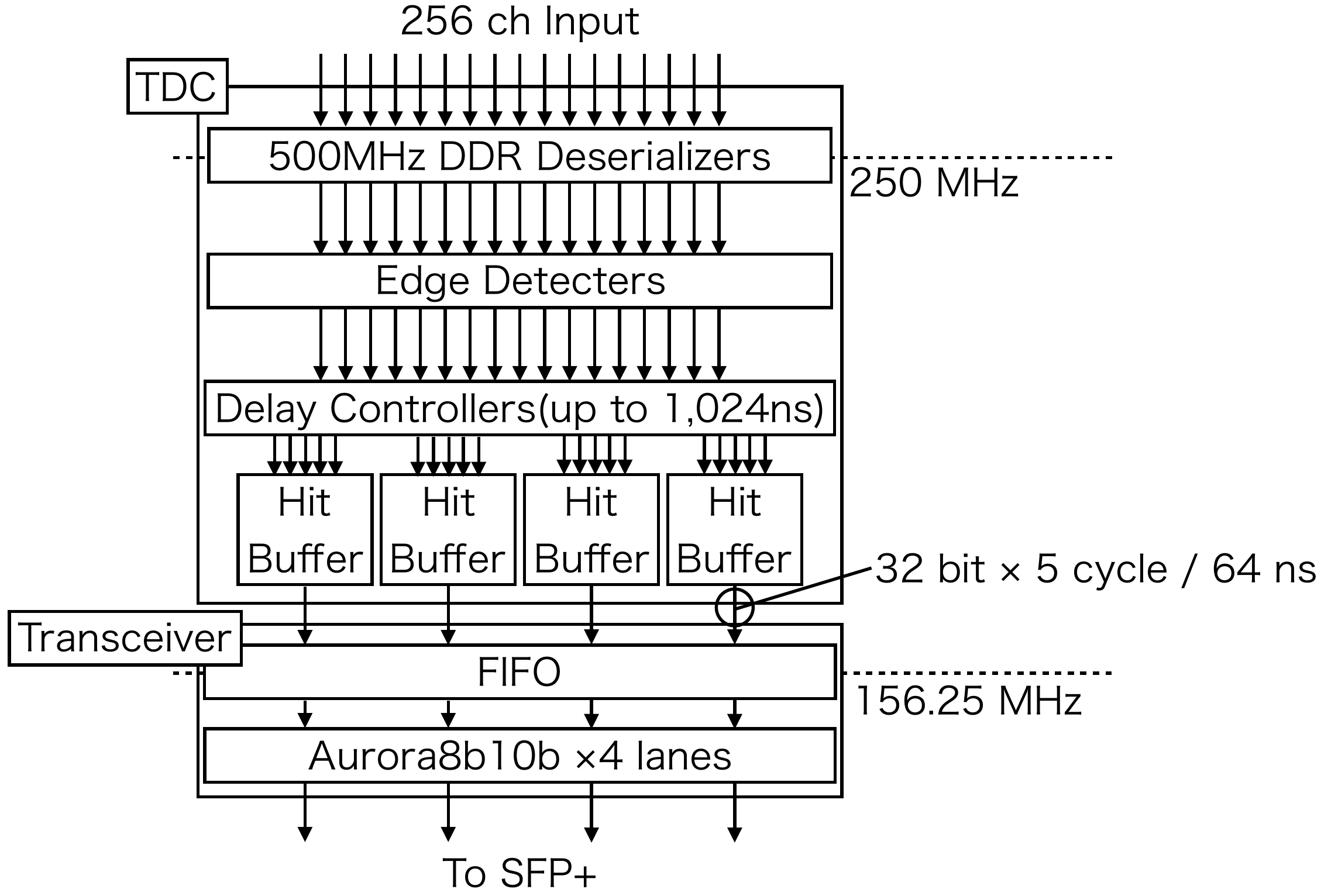}
\caption{The diagram of the firmware implemented in the TRG-MRG.}
\label{firmware-fig}
\end{figure}

The diagram of the firmware implemented in the FPGA is shown in Fig. \ref{firmware-fig}.
The firmware of TDC and transceiver sections is explained in the following paragraphs.
\subsubsection{TDC Section}
\label{firmware-TDC}
The TDC section consists of deserializers, edge detectors, delay controllers, and hit buffers.
The input signals are sampled with 1\,ns by 500\,MHz DDR (Double Data Rate) deserializers of Vivado IP core, ISERDESE2.
The component converts 1\,Gbps to four 250\,Mbps.
In the edge detector, leading-edges are detected from each 4\,bits data.
To calibrate the intrinsic time difference among channels, delays are added in the delay controller.
The component is implemented by using RAM based shift register to save the number of flip-flops and able to delay the data of each channels up to 1,024\,ns with a 4\,ns unit.
If the leading-edges are detected, the hit timing and channel number data are buffered in the hit buffer.
Maximal eight hits data are buffered for 64\,ns in each 64 channels.
The efficiency of event transfer with this criteria is discussed in Sec.\ref{transfer-eff}.
The data to the transceiver section have the width of 32\,bits and are output during 5\,cycles in each 64 channels.
\subsubsection{Transceiver Section}
The transceiver section consists of FIFO and Aurora8B/10B protocol\cite{Aurora}.
The Aurora8B/10B is a link-layer protocol for high-speed serial communication.
Because the clock frequency in the protocol is determined from the line rate and lane width of the transceiver, FIFO is installed for clock domain crossing.
In the Aurora transmitting, the 32\,bit/lane data are encoded to 40\,bit/lane data and serialized.
The data are deserialized and decoded at the stage of data receiving.
\section{Performance Evaluation}
Items of evaluated performance are described below.
\begin{itemize}
\item[$A$]Time resolution
\item[$B$]Integral non linearity (INL)
\item[$C$]Differential non linearity (DNL)
\item[$D$]Minimum pulse width
\item[$E$]Double pulse separation
\item[$F$]Latency
\item[$G$]Transfer efficiency
\end{itemize}
\subsection{Time Resolution}
\label{resolution}
\begin{figure}
\centering
\includegraphics[clip,width=8.5cm]{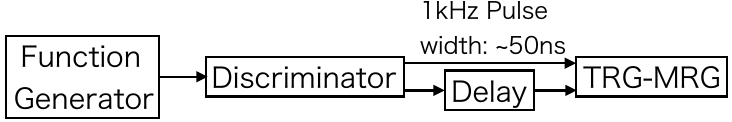}
\caption{The schematics of the circuit setup for the measurement of the time resolution.}
\label{resolution-circuit}
\end{figure}

\begin{figure}
\centering
\includegraphics[clip,width=8.5cm]{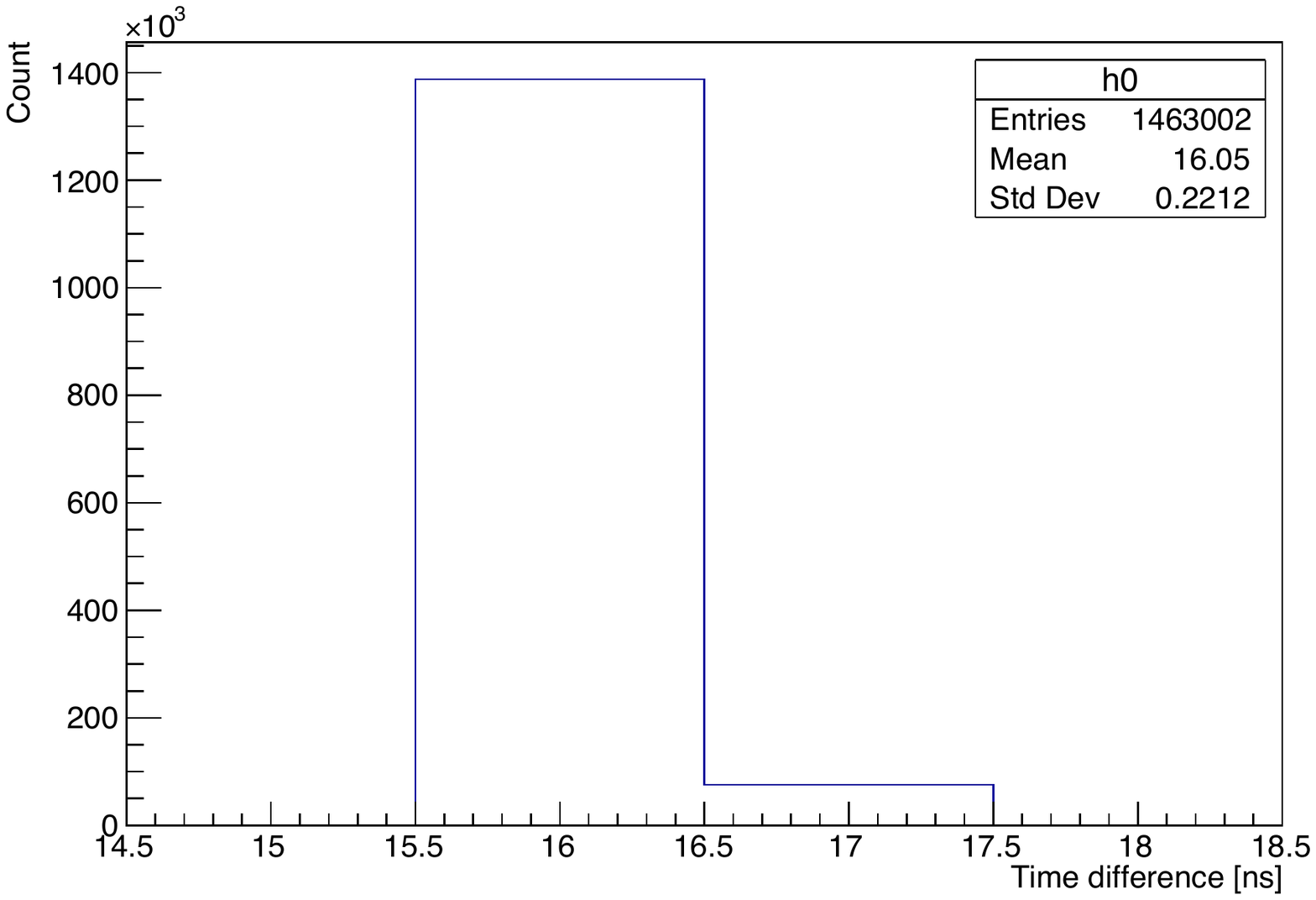}
\caption{An example of the distribution of measured time difference.}
\label{resolution-exp}
\end{figure}

\begin{figure}
\centering
\includegraphics[clip,width=8.5cm]{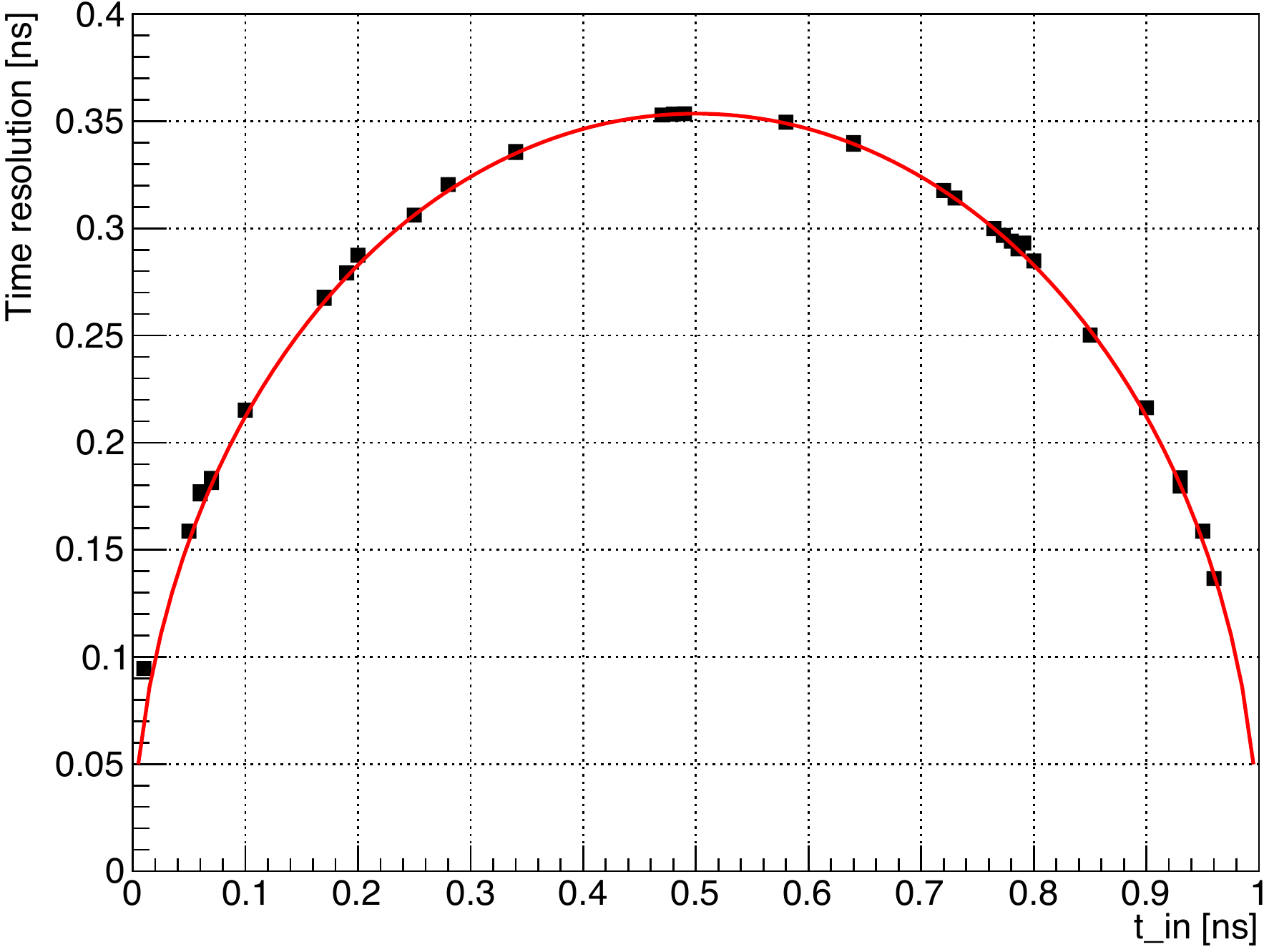}
\caption{The $t_{\rm{in}}$ dependency of the time resolution.
The black points indicate the measured value and the red line is the calculated value of quantization error.}
\label{resolution-result}
\end{figure}

The time resolution was evaluated by inputting two signals with a fixed delay to two channels of the TRG-MRG as illustrated in the Fig. \ref{resolution-circuit}.
The time difference of the output from the TRG-MRG was measured as shown in Fig. \ref{resolution-exp}.
The time resolution is defined from the distribution as $\sigma / \sqrt{2}$, $\sigma$ is defined as the standard deviation of the distribution.
Even if the TDC has no clock jitter, the time differences distribute at least 2\,LSB (Least Significant Bit), which is called as a quantization error.
The quantization error depends on the remainder of (time difference) / LSB, represented as $t_{\rm{in}}$ in this paper, as $\sqrt{t_{\rm{in}}(1 - t_{\rm{in}})}$.
The measured distribution is in good agreement with the expected quantization error as shown in Fig. \ref{resolution-result}.
The time resolution of better than 0.35\,ns are obtained.
\subsection{Integral Non Linearity}
\begin{figure}
\centering
\includegraphics[clip,width=8.5cm]{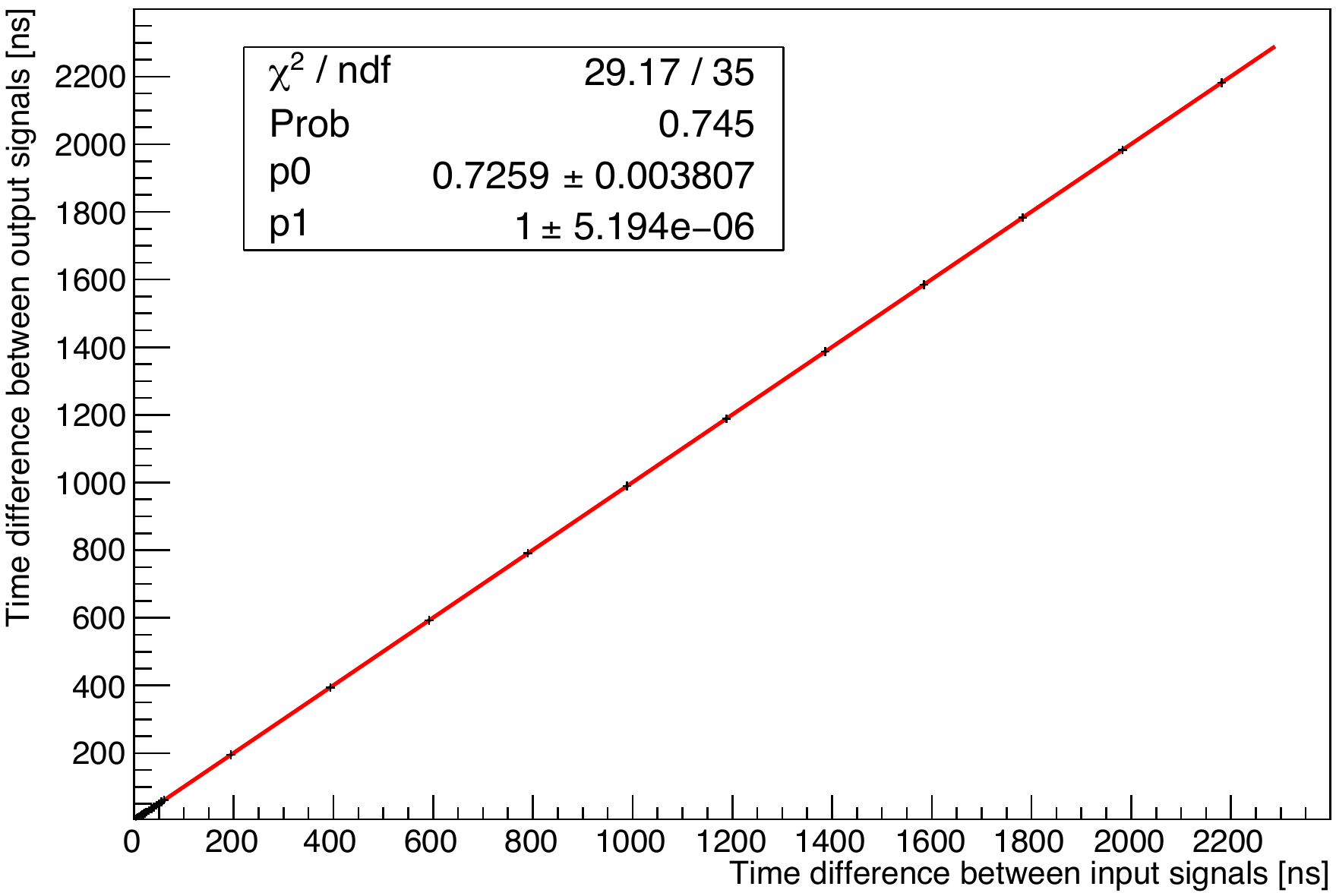}
\caption{The relation between input time difference and output time difference.
The black points indicate the measured value and the red line is the result of fitting.}
\label{inl-meas}
\end{figure}

\begin{figure}
\centering
\includegraphics[clip,width=8.5cm]{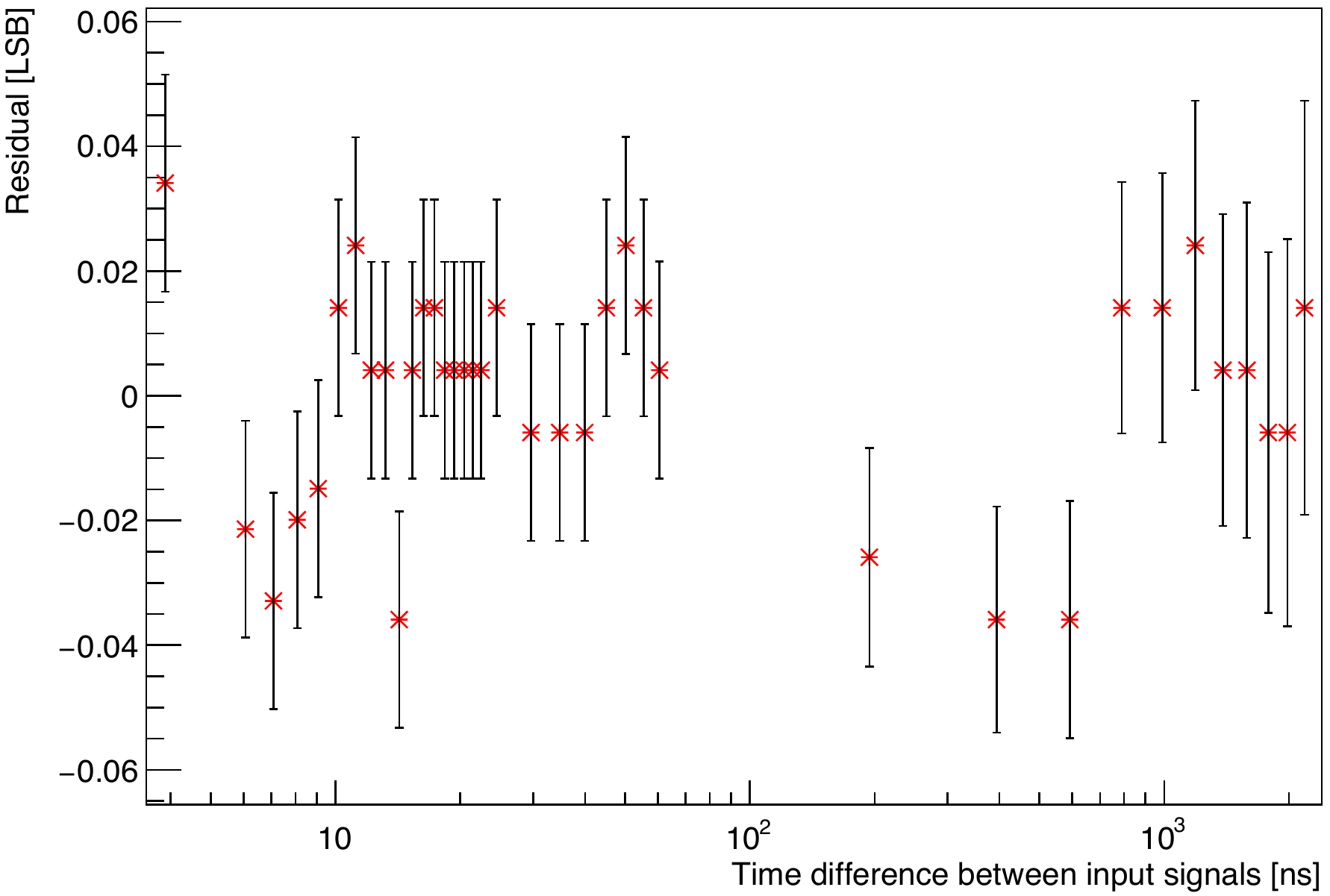}
\caption{The result of the INL measurement.}
\label{inl-res}
\end{figure}

The integral non linearity (INL) was estimated by the same data described in Sec. \ref{resolution}.
Figure \ref{inl-meas} shows the relation between input time difference and output time difference.
By fitting the measured points as A$t +$ B and calculating the residual between the measured points and the fitting line, the INL was estimated as the maximal value of the residual of [$-$0.04\,LSB, +0.04\,LSB] (fig. \ref{inl-res}).
The effect of INL turned out to be negligible for the performance.
\subsection{Differential Non Linearity}
\begin{figure}
\centering
\includegraphics[clip,width=8.5cm]{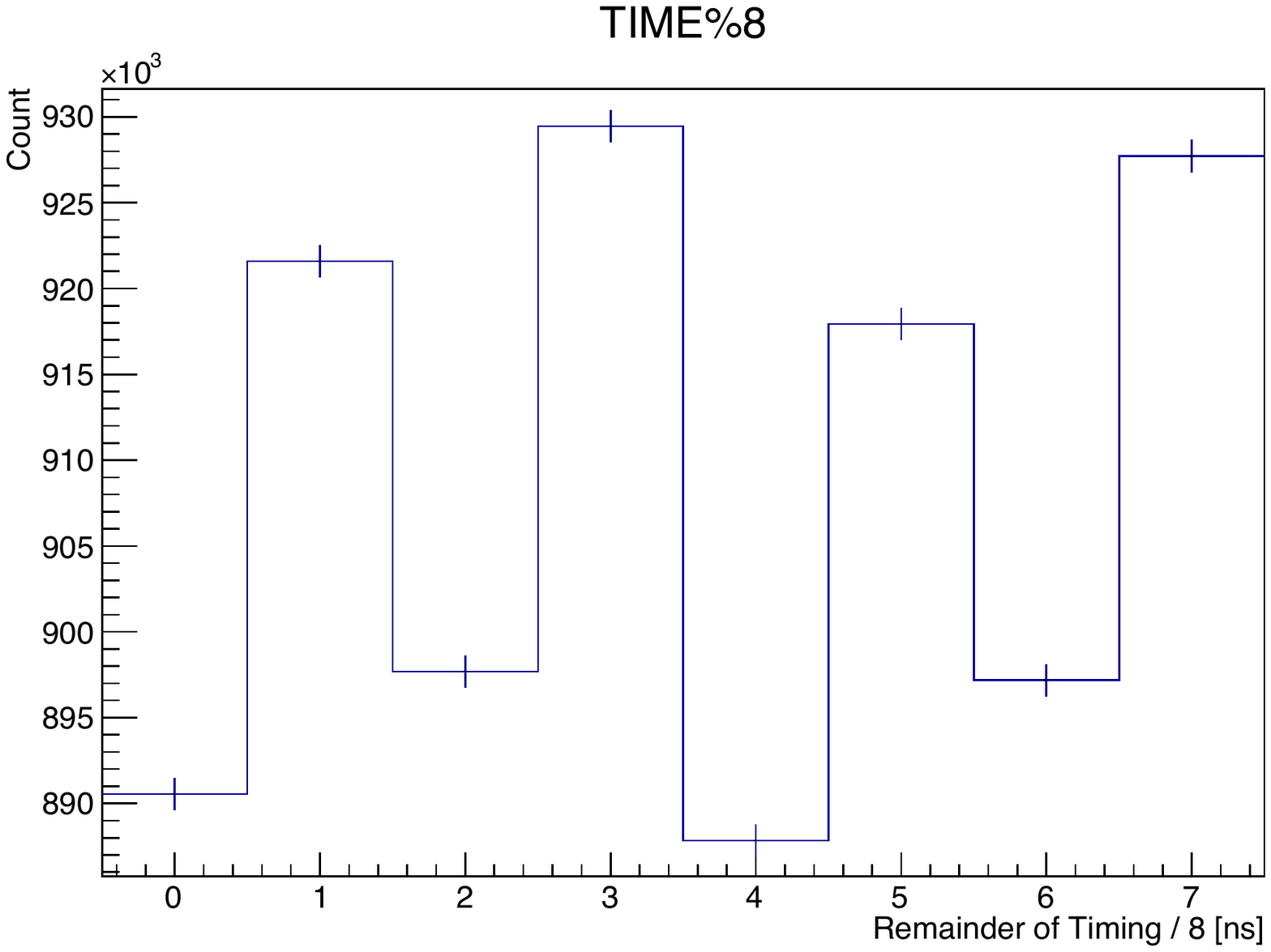}
\caption{The distribution of the edges of the 80.008\,ns clock.}
\label{dnl-dist}
\end{figure}

\begin{figure}
\centering
\includegraphics[clip,width=8.5cm]{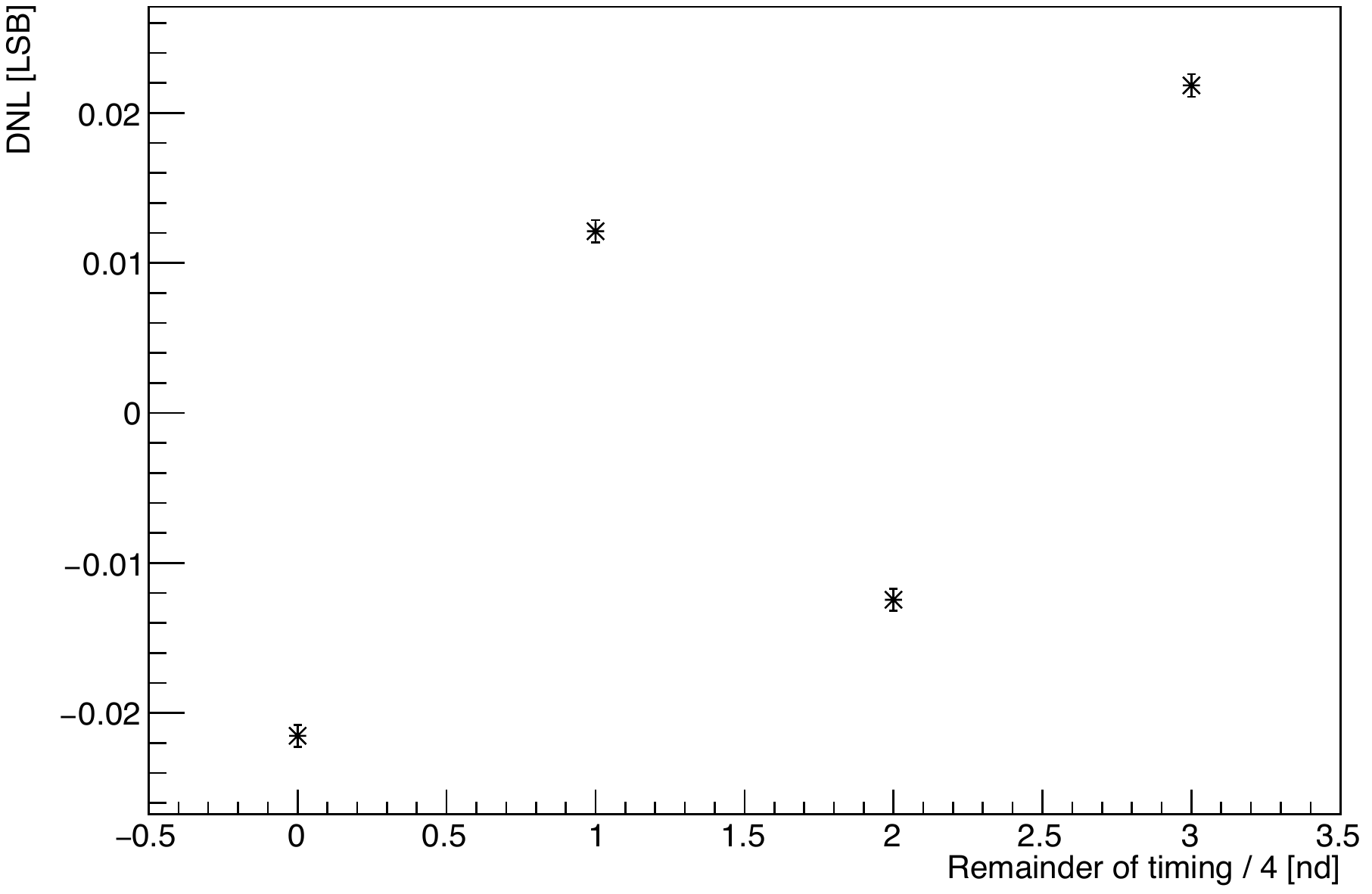}
\caption{The result of DNL measurement.}
\label{dnl-result}
\end{figure}

In the TRG-MRG, the differential non linearity (DNL) is expected to be originated from the deserializer.
The accuracy of the output from clock generator and the skew of interconnection length in the deserializer make the DNL worse.
As mentioned above, the input data are deserialized to 4\,bits in each 4\,ns.
Therefore, the DNL is expected to have a periodicity of 4\,ns.
The DNL measurement was performed by code density test with a clock with the period of 80.008\,ns.
The edges of the input clock is expected to distribute with the interval of 0.008\,ns into the expected periodicity of 4\,ns.
The distribution of the those edges is shown in Fig. \ref{dnl-dist}.
As expected, the 4\,ns periodicity is seen.
The DNL was estimated at [$-$0.022\,LSB, +0.022\,LSB], as shown in Fig. \ref{dnl-result}.
The effect of DNL turned out to be negligible for the performance.
\subsection{Minimum Pulse Width}
As mentioned previously, the TRG-MRG must detect the narrow signal of 3\,ns, expected in the experiment.
By inputting such narrow signals, the detection efficiency was measured.
As the result, it is understood that the TRG-MRG can detect signals of 1.0\,ns width in 100\% efficiency.
Therefore, the performance of the TRG-MRG about minimum pulse width satisfies the requirement from the experiment.
\subsection{Double Pulse Separation}
\begin{figure}
\centering
\includegraphics[clip,width=8.5cm]{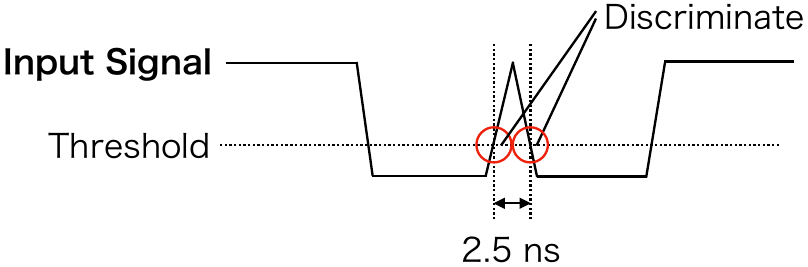}
\caption{A sketch of the input signals. The signals are shown as single end and negative logic for the simplification.}
\label{dps-sig}
\end{figure}

The double pulse separation was estimated by measuring the signal detection efficiency with changing the width between the trailing-edge of first signal and the leading-edge of second signal, shown in Fig. \ref{dps-sig}.
The TRG-MRG can discriminate two signals with the interval of 2.5\,ns in 100\% efficiency.
From the result, the inefficiency due to the TRG-MRG is estimated to be 0.27\% at the worst case.
\subsection{Latency}
The latency was evaluated in two sections, TDC and transceiver sections as defined in Sec.\ref{firmware}, separately.
\subsubsection{TDC Section}
\begin{figure}
\centering
\includegraphics[clip,width=8.5cm]{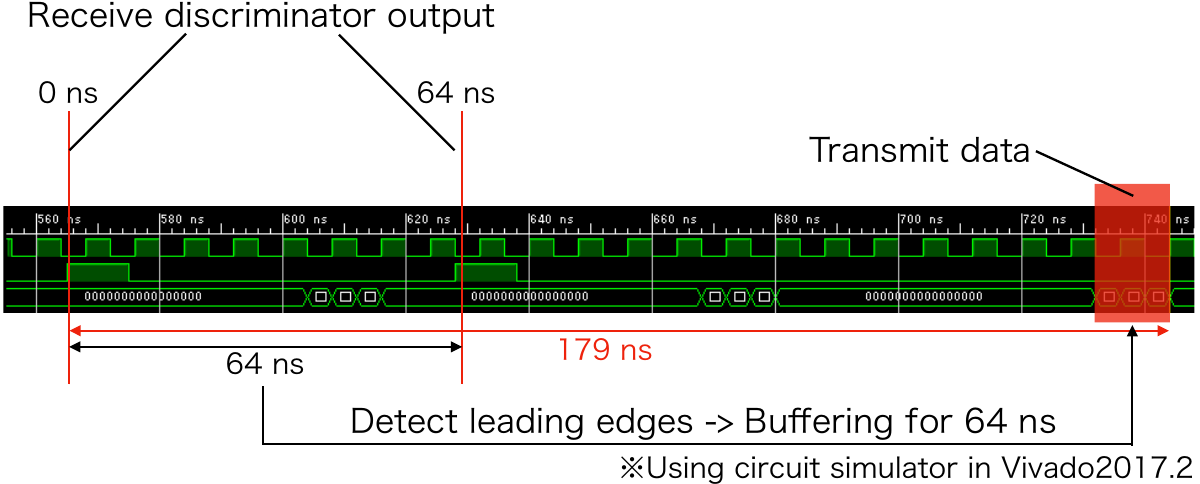}
\caption{The result of the estimation of the latency of the TDC section.}
\label{latency-TDC}
\end{figure}

The latency of the TDC section was estimated by using a logic simulator in Vivado.
The latency before the TRG-MRG is 600\,ns, which delays added by delay controller is included.
The result is shown in Fig. \ref{latency-TDC}.
Including the buffering-time of 64\,ns, the latency of the TDC section is maximal 179\,ns.
\subsubsection{Transceiver Section}
\begin{figure}
\centering
\includegraphics[clip,width=8.5cm]{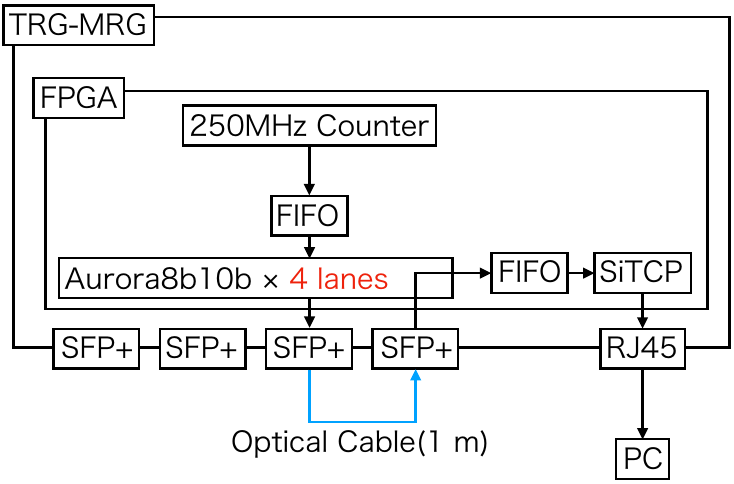}
\caption{The schematics of the circuit for the measurement of the latency of the transceiver section. The RJ45 connector is attached to one SFP+ port.}
\label{latency-transceiver-circuit}
\end{figure}

The latency of the transceiver section is measured by inputting the output of 250\,MHz counter to the FIFO and receiving the data passing Aurora and optical cable of 1\,m (expected length).
The latency is mainly defined from necessary time for data receiving (deserializing and decoding).
Figure \ref{latency-transceiver-circuit} shows a schematic view of the measurement.
The data are transmitted in five cycles (20\,ns) continuously in each 16\,cycles (64\,ns) of 250\,MHz clock, corresponding to the transmission in the experiment.

\begin{figure}
\centering
\includegraphics[clip,width=8.5cm]{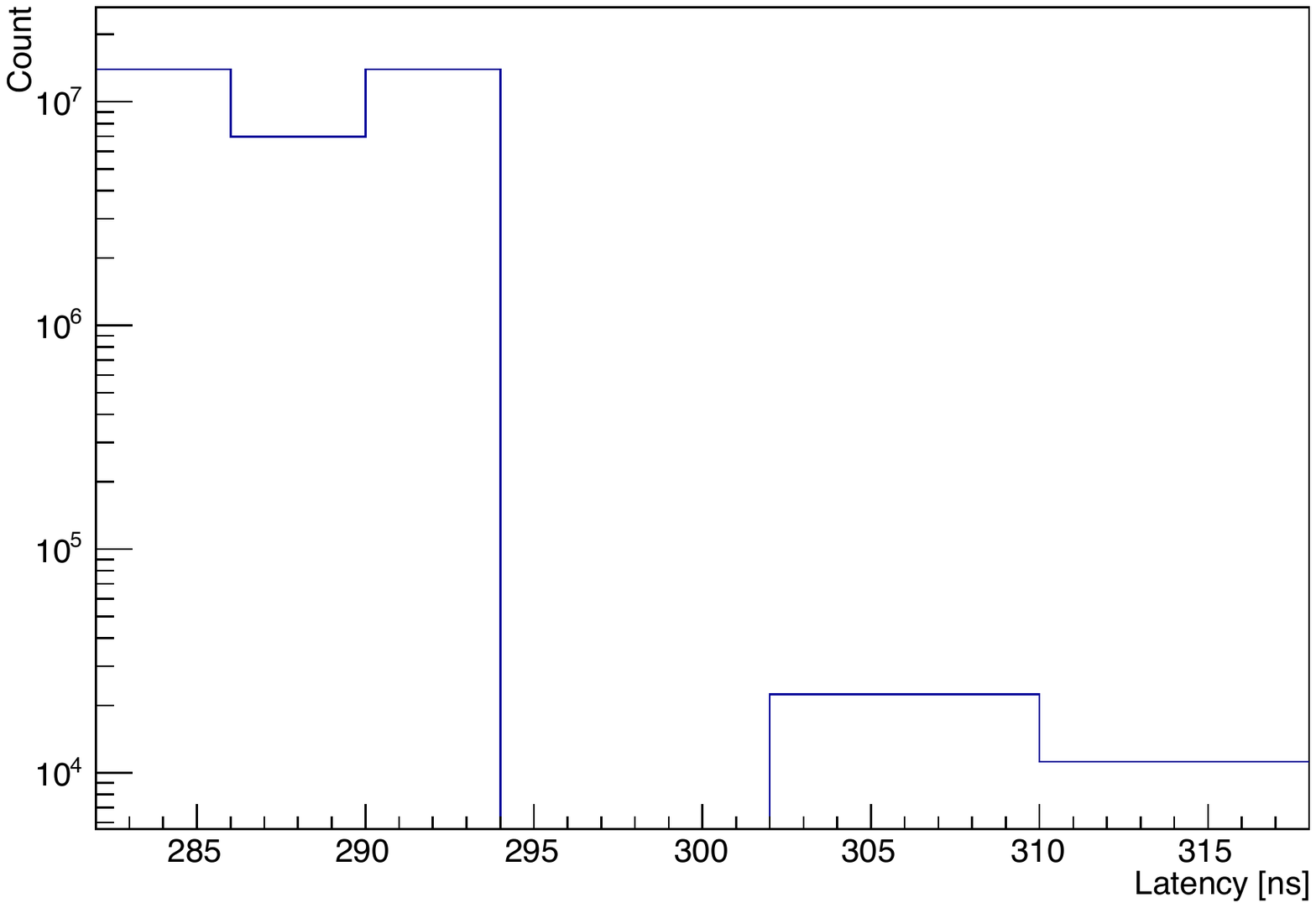}
\caption{The result of the measurement of the latency of the transceiver section.}
\label{latency-transceiver-result}
\end{figure}

The obtained result is shown in Fig. \ref{latency-transceiver-result}.
The latency of 290\,ns is obtained for 99.8\% data, which is consistent with the result from a measurement with a logic analyzer in Vivado.
On the other hand, the data of 0.2\% have longer latency of 310\,ns.
It seems to be due to the busy signal of Aurora protocol for clock compensation.
Clock compensation is the basic function of the Aurora protocol and outputs busy among three cycles (19.2\,ns) in each 2,500\,cycles of the clock for Aurora of 156.25\,MHz\cite{Aurora}.
The results of the value and percentage of the increase of the latency are consistent with the expectation from clock compensation.
As a result, the latency of the transceiver section is maximal 318\,ns.
\subsubsection{Total Latency}
The latency of the TDC and transmission is estimated to be maximal $179 + 318 = 497$\,ns.
It satisfies the requirement of less than 500\,ns.
\subsection{Transfer Efficiency}
\label{transfer-eff}
\begin{figure}
\centering
\includegraphics[clip,width=8.5cm]{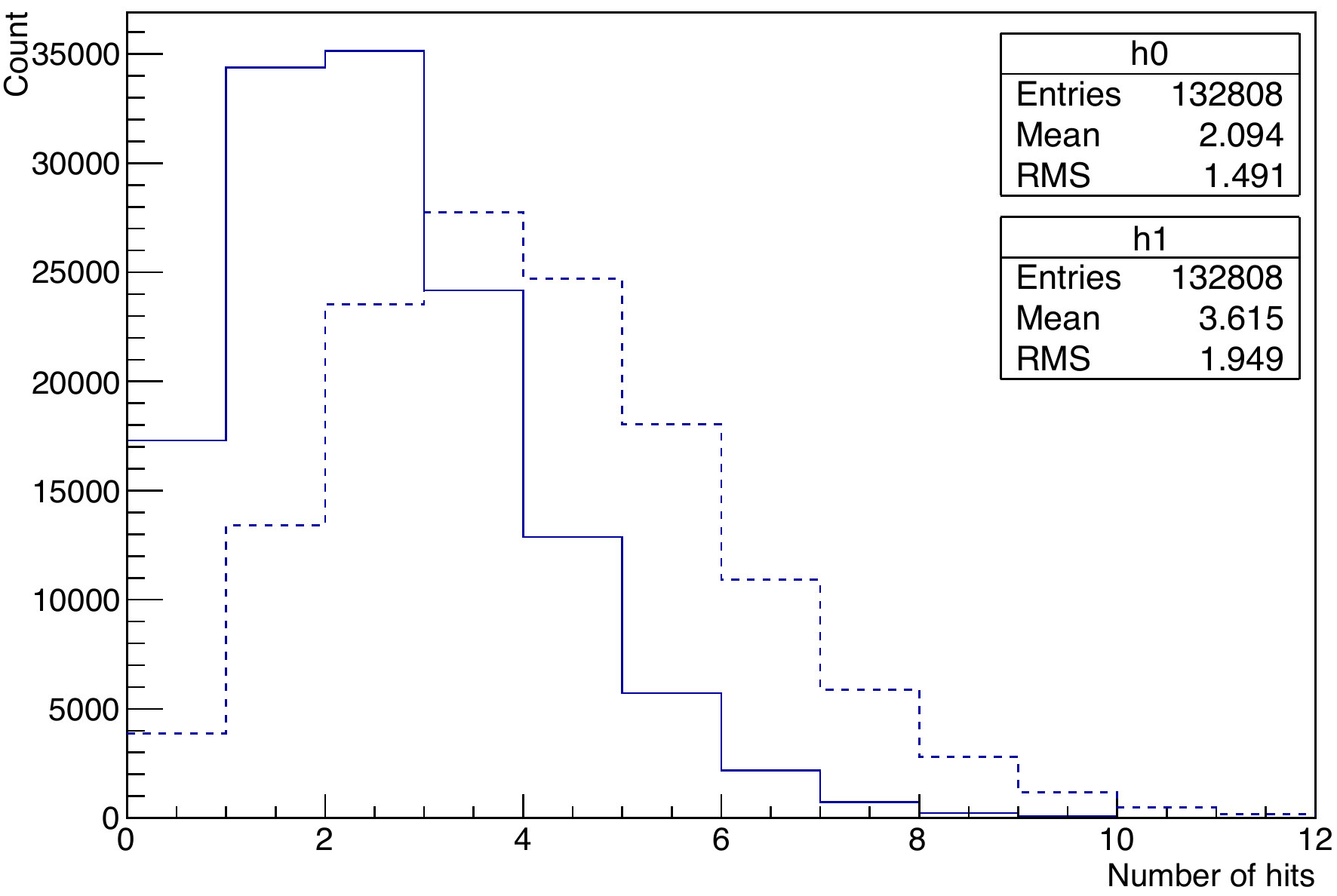}
\caption{Simulated hit multiplicities for 64\,ns with the beam intensity of $1\times10^{10}$\,/pulse (full line) and $2\times10^{10}$\,/pulse (dotted line).}
\label{eff-multi}
\end{figure}

\begin{figure}
\centering
\includegraphics[clip,width=8.5cm]{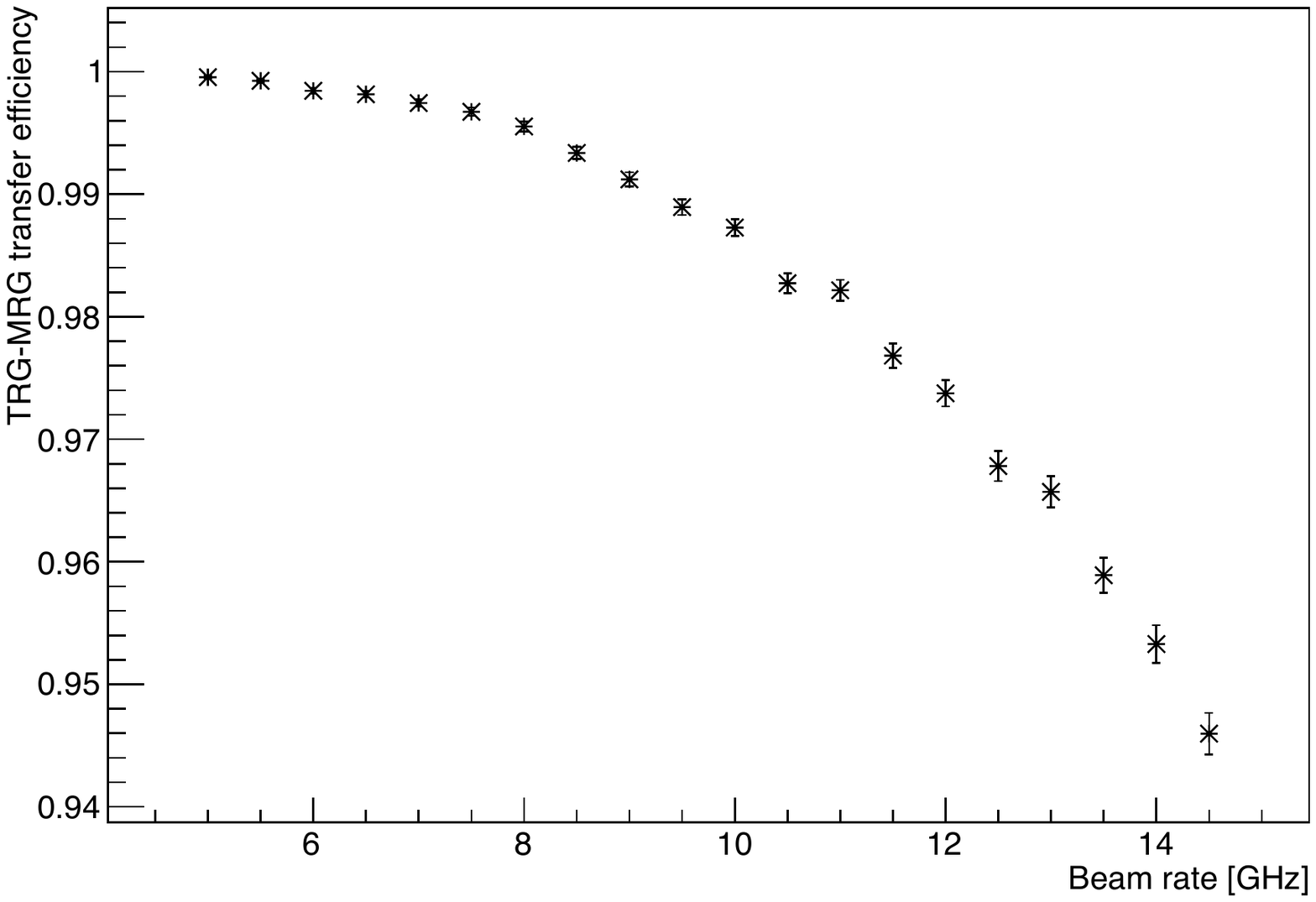}
\caption{The beam rate dependence of the transfer efficiency (simulation).}
\label{eff-result}
\end{figure}

As mentioned in Sec.\ref{firmware-TDC}, in the TRG-MRG, the hit data are transferred according to the criteria that the maximal eight hit data are buffered for 64\,ns in each 64 channels.
A detector simulation with a simulator of passage of particle, Geant4, was performed to estimate the transfer efficiency under the expected experimental condition\cite{Geant4-1, Geant4-2, Geant4-3}.
In the experiment, the proton-beam intensity is $1\times10^{10}$\,/pulse and the maximal single rate reachs 1\,MHz/ch.
For considering the micro structure of the beam intensity, instantaneous beam intensity distributes up to $2\times10^{10}$\,/pulse.
Figure \ref{eff-multi} shows the hit multiplicities for 64\,ns time windows under the beam intensity of $1\times10^{10}$\,/pulse (full line) and $2\times10^{10}$\,/pulse (dotted line).
The fraction of discarded hits depends on the beam intensity.
The beam rate dependence of the transfer efficiency is shown in Fig. \ref{eff-result}.
At the expected beam intensity of $1\times10^{10}$\,/pulse (5\,GHz), the transfer efficiency is expected to be 99.95\%.
Even at the intensity of $2\times10^{10}$\,/pulse (10\,GHz), the efficiency stays better than 98\%.
In conclusion, it is confirmed that he developed module meets the required performance.
\subsection{Summary of the Performance Evaluation}
At the last of this section, the result of the performance evaluation is summarized in TABLE \ref{summary}.
\begin{table}
\begin{center}
\caption{The result of the performance evaluation.}
\label{summary}
\begin{tabular}{|l||l|}
\hline
Point&Value \\ \hhline{|=#=|}
Time Resolution&$<$0.35\,ns \\ \hline 
INL&[$-$0.04\,LSB, +0.04\,LSB] \\ \hline
DNL&[$-$0.022\,LSB, +0.022\,LSB] \\ \hline
Minimum Pulse Width&100\% efficiency of 1.0\,ns width \\ \hline
Double Pulse Separation&100\% efficiency of 2.5\,ns interval \\ \hline
Latency&$<$497\,ns \\ \hline
Transfer efficiency&99.95\% at the expected condition \\ \hline
\end{tabular}
\end{center}
\end{table}
All the results satisfy the requirement of the experiment.
\section{Conclusion}
The J-PARC E16 experiment is planned in order to investigate the partial restoration of breaking of chiral symmetry at nuclear density.
To handle a massive number of trigger channels of 2,620, the trigger merging module, named TRG-MRG, has been developed.
The TRG-MRG consists of one main board and two mezzanine cards and will be installed between discriminators and the trigger decision module.
It works as 1\,ns TDC and data multiplexer with four 6.25\,Gbps transceivers.
From the results of the performance tests, for example time resolution, latency, transfer efficiency, it is confirmed that the TRG-MRG achieves the requirement for the experiment which will be started in JFY 2019.
\ifCLASSOPTIONcaptionsoff
  \newpage
\fi


\begin{thebibliography}{1}
\bibitem{E16-1}Y. Komatsu, \it et al.\rm, JPS Conf. Proc. 13 020005 (2017)
\bibitem{E16-2}Y. Morino, \it et al.\rm, JPS Conf. Proc. 8 022009 (2015)
\bibitem{GTR}Y. Komatsu, \it et al.\rm, Nucl. Instr. and Meth. A 732 (2013) 241
\bibitem{HBD}K. Aoki, \it et al.\rm, Nucl. Instr. and Meth. A 628 (2011) 300
\bibitem{APV}APV25-S1 User Guide Version 2.2, [online] available: \url{https://cds.cern.ch/record/1069892/files/cer-002725643.pdf}
\bibitem{DRS4}DRS4 datasheet rev. 0.9, [online] available: \url{https://www.psi.ch/drs/DocumentationEN/DRS4_rev09.pdf}
\bibitem{circuit-total}T. N. Takahashi, \it et al.\rm, J. Phys.: Conf. Ser. 664 082053 (2015)
\bibitem{DRS4-module}Open-it R\&D Project, ADC HRTDC with DRS4[online] available: \url{http://openit.kek.jp/project/DRS4ADC/public/drs4adc-public}
\bibitem{GTR-ASD}Y. Obara, \it et al.\rm, J. Phys.: Conf. Ser. 664 082043 (2015)
\bibitem{HBD-ASD}Open-it R\&D Project, HBD trigger ASIC[online] available: \url{http://openit.kek.jp/project/e16_hbd_trigger/e16_hbd_trigger}
\bibitem{UT3}Belle I\hspace{-1pt}I Technical Design Report, [online] available: \url{https://arxiv.org/pdf/1011.0352.pdf}
\bibitem{FTSW}M. Nakao, 2012 JINST7 C01028
\bibitem{kintex7}7 Series FPGAs Data Sheet: Overview, [online] available: \url{https://japan.xilinx.com/support/documentation/data_sheets/ds180_7Series_Overview.pdf}
\bibitem{spaltan3}Spartan-3AN FPGA Family Data Sheet, [online] available: \url{https://www.xilinx.com/support/documentation/data_sheets/ds557.pdf}
\bibitem{Aurora}Aurora 8B/10B v11.1 LogiCORE IP Product Guide, [online] available: \url{https://japan.xilinx.com/support/documentation/ip_documentation/aurora_8b10b/v11_1/pg046-aurora-8b10b.pdf}
\bibitem{Geant4-1}S. Agostinelli, \it et al.\rm, Nucl. Instr. and Meth. A 506 (2003) 250
\bibitem{Geant4-2}J. Allison, \it et al.\rm, IEEE Trans. Nucl. Sci. 53 (2006) 270
\bibitem{Geant4-3}J. Allison, \it et al.\rm, Nucl. Instr. and Meth. A 835 (2016) 186
\end{thebibliography}
\end{document}